\documentclass[12pt,a4paper]{article}
\usepackage{epsfig}
\begin{document}
\title{The littlest Higgs model with T-parity and single top production at $ep$ collision}
\author{Chong-Xing Yue, Jia Wen, Jin-Yan Liu, Wei Liu\\
{\small Department of Physics, Liaoning  Normal University, Dalian
116029, P.R.China}
\thanks{E-mail:cxyue@lnnu.edu.cn}}
\date{\today}

\maketitle

\begin{abstract}
\hspace{5mm}Based on calculating the contributions of the littlest
Higgs model with T-parity (called $LHT$ model) to the anomalous top
coupling $tq\gamma$ ($q=u$ or $c$), we consider single top
production via the t-channel partonic process $eq\rightarrow et$ at
$ep$ collision. Our numerical results show that the production cross
section in the $LHT$ model can be significantly enhanced relative to
that in the standard model.
\end {abstract}
\newpage

\noindent{\bf I. Introduction}

The top quark with a mass of the order of the electroweak scale
$m_{t}\sim172GeV$[1] is the heaviest particle yet discovered and
might be the first place in which the new physics effects could be
appeared. The correction effects of new physics on observables for
top quark are often important than for other fermions. In
particular, the anomalous top couplings, which affect top production
and decay at high energy colliders, offer a unique place for testing
the standard model ($SM$) flavor structure[2].

In the $SM$, the anomalous top quark couplings $tqV$ $(q=c$ or $u$
quark and $V=\gamma,Z$ or $g$ gauge bosons), which are arised from
the flavor changing ($FC$) interactions, vanish at tree level but
can be generated at the one-loop level. However, they are very
suppressed by the $GIM$ mechanism, which can not be detected in the
present and near future high energy experiments. It is well known
that the anomalous top quark couplings $tqV$ may be large in some
new physics models beyond the $SM$ and single top production is
sensitive to these types of couplings. Thus, studying the
contributions of the couplings $tqV$ to single production of the top
quark is of special interest. It will be helpful to test the $SM$
flavor structure and new physics beyond the $SM$.

The $ep$ collider, called $HERA$ collider with the center-of-mass
($c.m.$) energy $\sqrt{s}=320GeV$ or $THERA$ collider with the
$c.m.$ energy $\sqrt{s}=1TeV$[3], is the experimental facility where
high energy electron-proton and positron-proton interactions can be
studied. Within the $SM$, single top quark can not be produced at an
observable level in the $HERA$ and $THERA$ collider experiments[4].
However, the $HERA$ and $THERA$ colliders can provide a good
sensitivity on the couplings $tqV$ via single top production[5].
Some studies about this type of single top quark production have
appeared in the literature[6,7], which have shown that the $HERA$
and $THERA$ colliders are powerful tools for searching for the
anomalous top quark couplings $tqV$.

The littlest Higgs model with T-parity (called $LHT$ model)[8] is
one of the attractive little Higgs models. To simultaneously
implement T-parity, the $LHT$ model introduces new mirror fermions.
Under T-parity, particle fields are divided into T-even and T-odd
sectors. The T-even sector consists of the $SM$ particles and a
heavy top quark $T_{+}$, while the T-odd sector contains heavy gauge
bosons ($B_{H}, Z_{H},W_{H}^{\pm}$), a scalar triplet ($\Phi$), and
so-called mirror fermions. The flavor mixing in the mirror fermion
sector gives rise to a new source of flavor violation, which might
generate significantly contributions to some flavor violation
processes[9,10,11,12,13,14]. In this paper, we will concentrate our
attention on single top production via the t-channel partonic
process $eq\rightarrow et$ at the $HERA$ and $THERA$ collider
experiments.

In the following section, we will give our numerical results in
detail. Our conclusion and a simple discussion are given in section
III.

\noindent{\bf II. Numerical results}

At the $HERA$ and $THERA$ colliders, single top quark can be
produced via the charged current ($CC$) process and the neutral
current ($NC$) process. In the $SM$, the former process can proceed
at the tree level, and its cross section is less than $1fb$[15],
while the latter process can only proceed at the one-loop level,
which is $GIM$ suppressed. However, the $NC$ process $ep\rightarrow
et+X$ is sensitive to the anomalous top quark couplings $tqV$.

The t-channel partonic process $eq\rightarrow et$ for the $NC$
process $ep\rightarrow et+X$ can obtain contributions from the
anomalous top quark couplings $tq\gamma$ and $tqZ$ via $\gamma$
exchange and $Z$ exchange, respectively. However, the contribution
from $Z$ exchange is several orders of magnitude smaller than that
from $\gamma$ exchange. Thus, we will neglect the contributions of
the coupling $tqZ$ in our numerical estimation. Then, in the context
of the $LHT$ model, the differential cross section of the partonic
process $e(P_{e})+q(P_{q})\rightarrow e(P_{e'})+t(P_{t})$ can be
written as[5,6]:
\begin{equation}
\frac{d\hat{\sigma}(\hat{s})}{dt}=\frac{(K_{tq\gamma})^{2}e^{4}}{32\pi
\hat{s}^{2}}[\frac{-2\hat{s}^{2}+2\hat{s}m_{t}^{2}-m_{t}^{4}}{t}+2m_{t}^{2}-2\hat{s}],
\end{equation}
where $t=q^{2}=(P_{e'}-P_{e})^{2}$ and $\sqrt{\hat{s}}$ is the
$c.m.$ energy of the t-channel partonic process $eq \rightarrow et$.
$K_{tq\gamma}$ is the effective coupling constant of the $tq\gamma$
vertex contributed by the $LHT$ model, in which we have drawn out
one factor $e$. In above equation, we have neglected the masses of
the incoming quarks and the electrons. However, for the Mandelstam
variable $t$, to avoid divergence we will use $m_{e}=0.511MeV$ for
its upper and lower limits in the phase space integral.

\begin{figure}[htb]
\vspace{-5cm}
\begin{center}
\epsfig{file=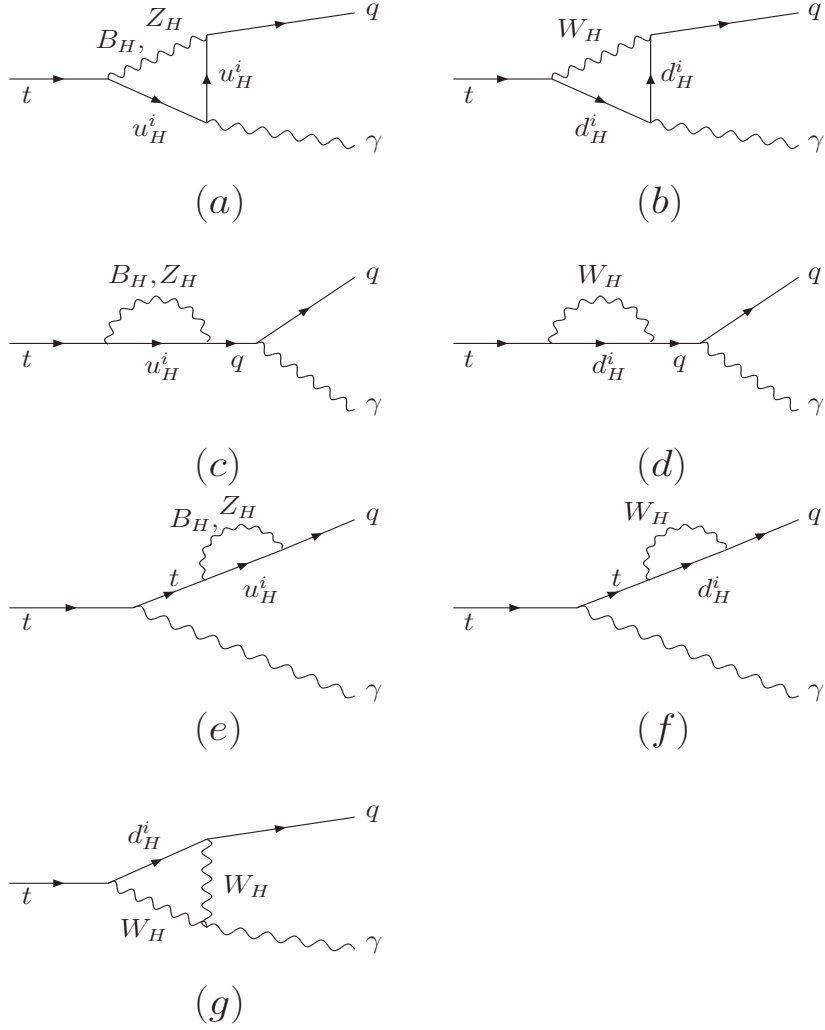,width=550pt,height=700pt} \vspace{-10.5cm}
\hspace{1cm} \vspace{3.6cm}\caption{ Feynman diagrams for the
$tq\gamma$ $(q=u$, or $c)$ vertex in the $LHT$ model.}
\end{center}
\end{figure}
\vspace{-0.5cm}

The effective cross section $\sigma(s)$ of single top production via
the partonic process $eq\rightarrow et$ at $ep$ collision can be
obtained by folding the cross section $\hat{\sigma}(\hat{s})$ with
the parton distribution function ($PDF$):
\begin{equation}
\sigma(s)=\sum_{q=u,c}\int^{1}_{x_{min}}f_{q}(x,\mu)dx\int^{t_{max}}_{t_{min}}
\frac{d\hat{\sigma}(\hat{s})}{dt}dt
\end{equation}
with $x_{min}=\frac{m_{t}^{2}+m_{e}^{2}}{s}$ and $\hat{s}=xs$, in
which the $c.m.$ energy $\sqrt{s}$ is taken as $320GeV$ for the
$HERA$ collider and as $1TeV$ for the $THERA$ collider. In our
numerical calculation, we will use $CTEQ6L$ $PDF$s[16] for the quark
distribution functions and assume that the factorization scale $\mu$
is of order $m_{t}$. The upper and lower limits of the Mandelstam
variable $t$ are taken as:
\begin{equation}
t_{max}\simeq -\frac{m_{e}m_{t}^{2}}{\widehat{s}},
\hspace{2.5cm}t_{min}\simeq
m_{t}^{2}-\widehat{s}-\frac{m_{e}^{2}(m_{e}^{2}-m_{t}^{2})}{\widehat{s}}.
\end{equation}

In the $LHT$ model, the anomalous top quark coupling $tq\gamma$ can
be induced by the interactions between the $SM$ quarks and the
mirror quarks mediated by T-odd gauge bosons ($B_{H},
Z_{H},W_{H}^{\pm}$), as shown in $Fig.1$. The heavy scalar triplet
$\Phi$ has not contributions to the coupling $tq\gamma$ at the order
of $\nu^{2}/f^{2}$[10,11]. So, in our numerical estimation, we will
neglect its contributions.

Using the relevant Feynman rules given in Refs.[10,17], we can
calculate the contributions of the $LHT$ model to the coupling
constant $K_{tq\gamma}$ of the anomalous top quark coupling vertex
$tq\gamma$. [Using the value of $K_{tq\gamma}$ for $q=c$, we have
calculated the branching ratio $Br(t\rightarrow c\gamma)$ and find
that our numerical result is approximately consistent with that
given by Ref.[13]]. Observably, the cross section $\sigma(s)$ of
single top production at $ep$ collision is dependant on the model
dependant parameters $f, M_{u^{i}_{H}}, M_{d^{i}_{H}}$, and
$(V_{Hu})_{ij}$. The matrix elements $(V_{Hu})_{ij}$ can be
determined via $V_{Hu}=V_{Hd}V_{CKM}^{+}$. The matrix $V_{Hd}$ can
be parameterized in terms of three mixing angles and three phases,
which can be probed by $FCNC$ processes in $K$ and $B$ meson
systems, as discussed in detail in Refs.[10,12]. It is convenient to
consider several representative scenarios for the structure of the
matrix $V_{Hd}$. To simplify our calculation, we concentrate our
study on the following two scenarios for the structure of the mixing
matrix $V_{Hd}$:

Case I: $V_{Hd}=I$,    $V_{Hu}=V_{CKM}^{+}$;

Case II: $s_{23}^{d}=\frac{1}{\sqrt{2}},
s_{12}^{d}=0,s_{13}^{d}=0,\delta_{12}^{d}=0,\delta_{23}^{d}=0,\delta_{13}^{d}=0$.

It has been shown that, in both above cases, the constraints on the
mass spectrum of the mirror fermions are very relaxed[10,12].
Furthermore, the masses of up and down type mirror quarks are equal
to each other at the leading order of $\nu/f$. Thus, we assume that
the masses of mirror quarks have the relations $
M_{u^{i}_{H}}=M_{d^{i}_{H}}=M_{1}(i=1,2)$ and $M_{u^{3}_{H}}=
M_{d^{3}_{H}}=M_{2}$. In our numerical estimation, we will take the
scale parameter $f$ and the mass parameters $M_{1}$, $M_{2}$ as free
parameters.
\begin{figure}[htb]
\vspace{-0.5cm}
\begin{center}
\epsfig{file=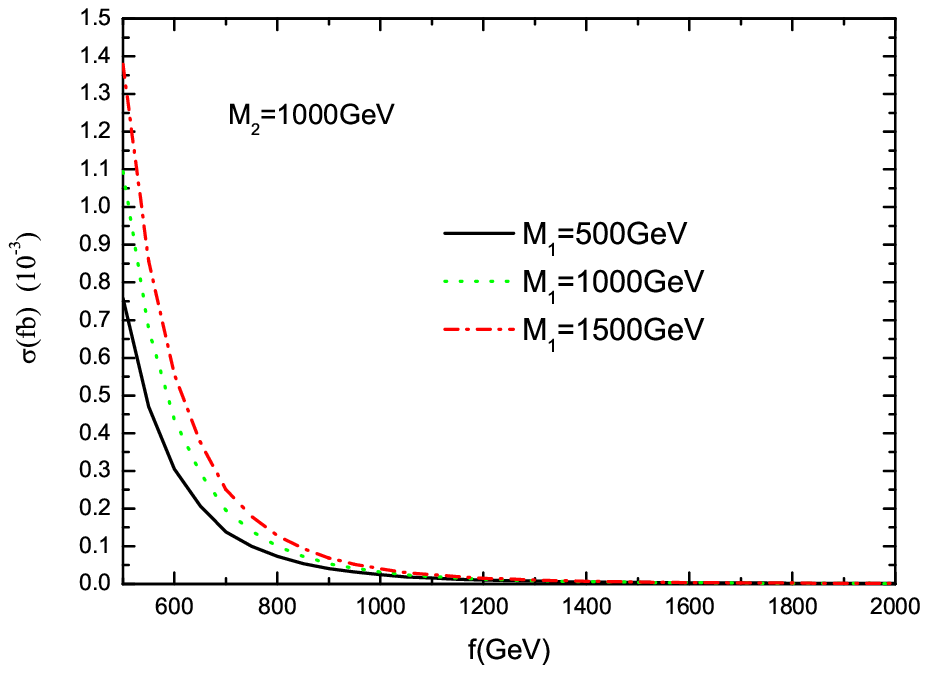,width=300pt,height=200pt}
\hspace{0cm}\vspace{-1cm}
\epsfig{file=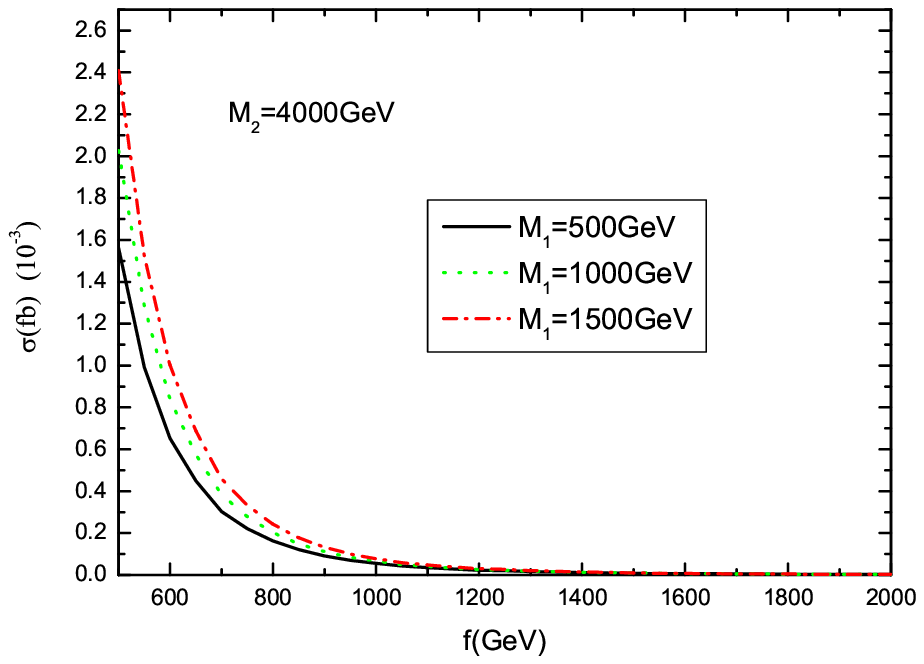,width=300pt,height=200pt} \hspace{0cm}
\vspace{-1cm}
 \caption{In Case I, the cross section $\sigma(s)$ of single top production at the $THERA$ collider
 \hspace*{1.8cm}with $\sqrt{s}=1TeV$ as a function of $f$ for different values of $M_{1}$ and $M_{2}$.}
 \label{ee}
\end{center}
\end{figure}

Our numerical results for Case I are summarized in $Fig.2$ and
$Fig.3$. In these two figures, we have taken the values of the $CKM$
matrix elements $(V_{CKM})_{ij}$ given by Ref.[18], in which
$V_{CKM}$ is constructed based on $PDG$ parameterization[19]. For
the $HERA$ collider with $\sqrt{s}=320GeV$, the value of $\sigma(s)$
is smaller than $1\times10^{-4}fb$ in most of the parameter space of
the $LHT$ model. So, in $Fig.2$, we only plot the production cross
section $\sigma(s)$ at the $THERA$ collider with $\sqrt{s}=1TeV$ as
a function of the scale parameter $f$ for different values of the
mass parameters $M_{1}$ and $M_{2}$. One can see from $Fig.2$ that
the value of the production cross section $\sigma(s)$ increases as
the scale parameter $f$ decreases. The value of $\sigma(s)$ is
insensitive to the mirror quark masses. The value of $\sigma(s)$ can
reach $2.2\times10^{-3}fb$ for $f=500GeV$, $M_{1}=1500GeV$ and
$M_{2}=4000GeV$,  which is too small to be detected at the $THERA$
collider with $\sqrt{s}=1TeV$. In $Fig.3$, we plot the production
cross section $\sigma(s)$ as function of the $c.m.$ energy
$\sqrt{s}$ for $M_{1}=1000GeV$, $M_{2}=3000GeV$, and three values of
the scale parameter $f$. From $Fig.3$, we can see that, even for
$f=500GeV$ and $\sqrt{s}=1.5TeV$, the value of $\sigma(s)$ is only
$3.7\times10^{-3}fb$.

\vspace{1cm}
\begin{figure}[htb]
\vspace{-0.5cm}
\begin{center}
\epsfig{file=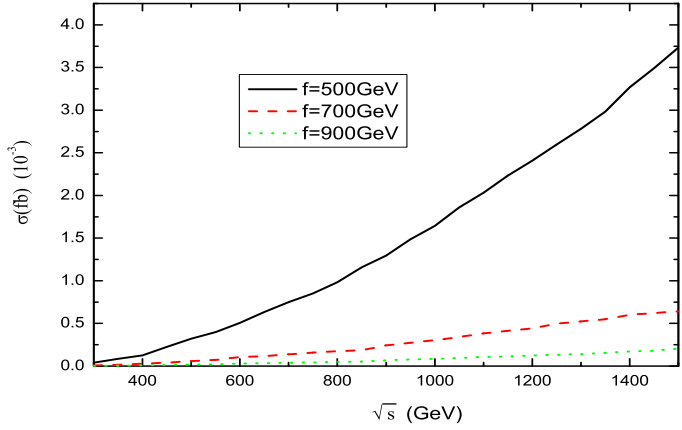,width=300pt,height=200pt} \vspace{-0cm}
\caption{In Case I, the cross section $\sigma(s)$ a function of the
$c.m.$ energy $\sqrt{s}$ for $M_{1}=\hspace*{1.8cm}1000GeV$,
$M_{2}=3000GeV$ and three values of $f$. }\label{ee}
\end{center}
\end{figure}

\begin{figure}[htb]
\vspace{-2.0cm}
\begin{center}
\epsfig{file=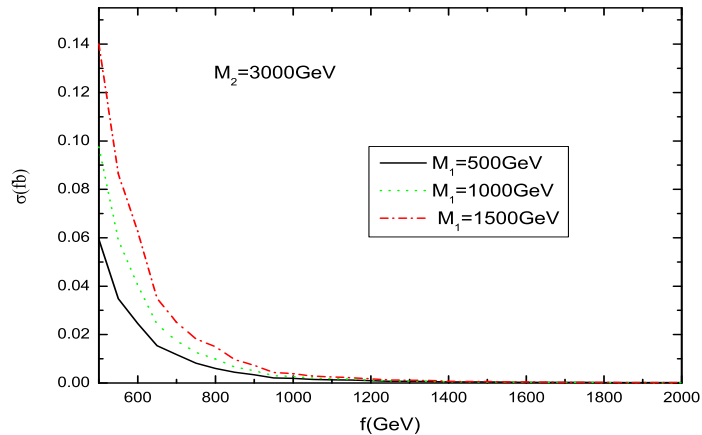,width=300pt,height=203pt}
 \caption{In Case II, the cross section $\sigma(s)$ at the $THERA$  collider with
 $\sqrt{s}=1TeV$ \hspace*{1.8cm}as a function of $f$ for $M_{2}=3000GeV$ and three  values of $M_{1}$.}
 \label{ee}
\end{center}
\end{figure}

For Case I we have assumed $V_{Hd}=I$ and $V_{Hu}=V_{CKM}^{+}$, for
Case II we will assume $V_{Hu}=V_{Hd}V_{CKM}^{+}$ and
\begin{equation}
V_{Hd}=\left(\begin{array}{ccc}1&0&0
\\0&\frac{1}{\sqrt{2}}&\frac{1}{\sqrt{2}}
\\0&-\frac{1}{\sqrt{2}}&\frac{1}{\sqrt{2}}
\end{array}\right).
\end{equation}

Observably, the value of the factor
$\lambda_{i}=(V_{Hu})_{ij}^{\ast}(V_{Hu})_{i3}$ for Case II is
different from that for Case I, which makes  the effective
production cross section $\sigma(s)$ of the t-channel partonic
process $eq\rightarrow et$ for Case II differ from that for Case I.
In $Fig.4$, the dependance of the cross section $\sigma(s)$ on the
scale parameter $f$ is presented for Case II. One can see from
$Fig.4$ that the cross section $\sigma(s)$ of single top production
for Case II is larger than that for Case I. For $f=500GeV$,
$M_{2}=3000GeV$, and $500GeV\leq M_{1}\leq 1500GeV$, the value of
$\sigma(s)$ is in the range of $5.9\times 10^{-2}fb\sim 0.14 fb$. In
$Fig.5$, we plot $\sigma(s)$ as a function of the $c.m.$ energy
$\sqrt{s}$ for $M_{1}=1000GeV$, $M_{2}=3000GeV$ and three values of
the scale parameter $f$. The value of $\sigma(s)$ can reach $0.18fb$
for $f=500GeV$ and $\sqrt{s}=1.5TeV$. If we assume that the $THERA$
collider with $\sqrt{s}=1TeV$ has a yearly integrated luminosity of
$\pounds=100fb^{-1}$, then there will be several
 tens of single top events to be generated.

\begin{figure}[htb] \vspace{-1.5cm}
\begin{center}
\epsfig{file=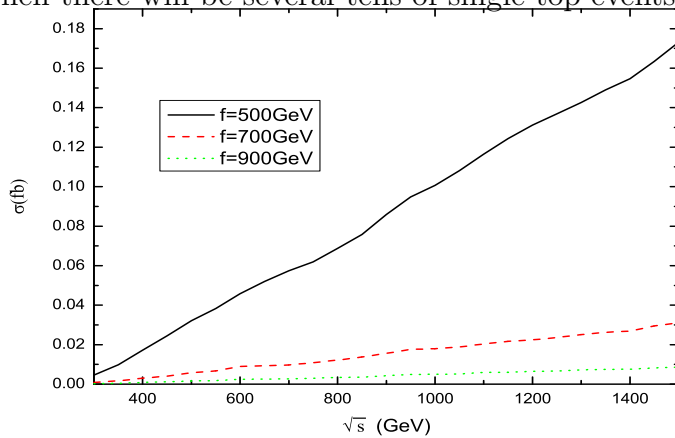,width=300pt,height=203pt}
 \hspace{0cm}\vspace{0cm}
 \caption{Same as $Fig.3$ but for Case II.}
 \label{ee}
\end{center}
\end{figure}

\noindent{\bf III. Conclusion and discussion}

 The $LHT$ model is one
of the attractive little Higgs models that is consistent with
electroweak precision tests but also has a much richer flavor
structure described by new flavor mixing matrices. This feature
makes that the $LHT$ model might generate significant contributions
to some flavor violation processes. Based on calculating the
contributions of the $LHT$ model to the anomalous top quark coupling
$tq\gamma$ ($q=u$ or $c$), single top quark production via the
t-channel partonic process $eq\rightarrow et$ at $ep$ collision is
considered in this paper. Our numerical results show that the
production cross section is too small to be measured at the $HERA$
collider experiments. However, with reasonable values of the
parameters in the $LHT$ model, the production cross section can
reach $0.14fb$ at the $THERA$ collider with $\sqrt{s}=1TeV$.
Certainly, the effects of the $LHT$ model on single top quark
production whether can be detected in future $THERA$ collider,
depending on its luminosity.

Single top quark production at $ep$ collision can also be induced by
the anomalous top quark coupling $tqg$. Using the constraint on the
single top production cross section obtained at the $HERA$ collider,
Ref.[20] has given upper limits on the coupling constant of the
anomalous coupling $tqg$. The $LHT$ model can generate a large
anomalous top quark coupling $tqg$[13], which can also produce
significant contributions to single top production at $ep$
collisions. However, compared single top production based on the
t-channel partonic process $eq\rightarrow et$, it has an additional
light jet from a gluon or up type quarks. Thus, single top
production at $ep$ collision induced by the anomalous top quark
coupling $tqg$ should be further studied.

\vspace{1.0cm}

\noindent{\bf Acknowledgments}

This work was supported in part by the National Natural Science
Foundation of China under Grants No.10675057 and Foundation of
Liaoning  Educational Committee(2007T086).

\end{document}